\documentstyle[aps,manuscript]{revtex}  

\newcommand{\tr}{{\tilde\rho}}
\newcommand{\I}{{\tilde I}}

\begin{document}               

\title{Diffusional Relaxation in Random Sequential Deposition}
\author{Eli Eisenberg$^{1,2}$ and Asher Baram$^{1}$}
\address{$^1$Soreq NRC, Yavne 81800, Israel \\
	$^2$Department of Physics, Bar-Ilan University, 
	Ramat-Gan 52900, Israel}

\maketitle
\begin{abstract}
The effect of diffusional relaxation on the random sequential 
deposition process is studied in the limit of fast deposition.
Expression for the coverage as a function of time are analytically 
derived for both the short-time and long-time regimes. These results
are tested and compared with numerical simulations. 
\end{abstract}

\bigskip

\leftline{{\bf PACS}: 05.70Ln, 68.10Jy, 82.20Wt}

Diffusional relaxation in irreversible deposition processes of
extended particles have been investigated recently
\cite{er,pr,rr,p,s1,s2}. In the
deposition process of immobile particles, terms random sequential
adsorbsion (RSA) the asymptotic behavior is dominated by the formation
of 'holes' too small for a new particle to fill in, resulting
in the jamming of the available area. Diffusion allows non-effective 
depositions to be corrected in latter stages of the process. 
Thus, the coverage reaches its maximal value - the closest packing 
value - for long times. The asymptotic filling process is dominated
by diffusive, power-law, approach to the steady state, as opposed to
the exponential convergence to the jamming limit in the immobile case
of lattice models.

Mainly, two kinds of relaxation mechanisms were studied. Firstly,
particle detachment has been studied, both experimentally and
analytically. Exact results were obtained for special values
of the parameters (detachment rate equal to deposition rate)
\cite{s1}.
In many experimental situations, another relaxation process, diffusion
of the deposited particles, is more significant. It has been shown
numerically that the effect of this diffusional relaxation process
in 1D is a $1/\sqrt{t}$ asymptotic approach to the closest packing 
value. This result was supported by analytical arguments as well
\cite{p}. A somewhat different model was also considered,
in which the absorbed dimers are allowed to dissociate into two
independent monomers. Each monomer can diffuse to its nearest 
neighbor sites. For this version of the model, for special values 
of the parameters (deposition rate twice 
the diffusion probability) exact solution is available \cite{s2}.

In this letter we study the combined effect of deposition and 
diffusion in 1D, in the regime in which separation of time scales is 
possible - the deposition rate is high, and may be considered 
infinite with respect to the diffusion process. This regime is highly
relevant for experimental interest. We find two series expansions, 
relevant for short and long times, which describe the whole dynamics 
of the filling process. 

Our model is defined as follows. We take an initially empty linear 
1D lattice containing $N$ sites, with periodic boundary 
conditions to minimize finite-size effects. First, particles are
deposited randomly on the lattice up to the jamming limit. Each 
deposited particle fills one lattice site, and excludes further 
deposition in his nearest neighbors sites (This is also equivalent 
to the deposition of dimers with no neighbor exclusion). Since the 
deposition is considered fast, this process takes zero time. Then, in 
each step one particle is selected, and with probability $\epsilon/2$ 
moves to the right or left, if possible. Each $N$ steps are 
considered a time step, and thus the probability for a particle to move
is $\epsilon$ per time unit. Practically, when simulating the model 
numerically, we can select at each time step only $\epsilon N$ 
particles
and move them with probability $1/2$ to the right or to the left
After every movement, if a space for another particle was formed it 
is immediately deposited. We then look at the density as a function 
of time.

In order to simplify the following analysis, we first set up our
terminology. At any stage of the process, the lattice is filled by
ordered regions, in which the particles are densely packed such that 
there is only one empty site between adjacent particles. In the 
border between these regions, there are two sucssesive empty sites
separating the areas. 
Each region is termed ``$k$-mer'' were $k$ is the number of particles
in the area. The initial concentration of the $k$-mers can be
easily calculated to be
\begin{equation}
\label{initial}
c_k^0= \frac{2e^{-4}}{1-e^{-2}}\left ( \frac{1-3e^{-2}}{1-e^{-2}}
\right )^{k-1}
\end{equation}
The only particles that move in any time are those which are at the 
edges of the $k$-mers. As a result of these motions the $k$-mers 
change their lengths. When a monomer moves, succesive three empty sites
are obtained, a new particle is deposited, and the $k$-mer and 
$k'$-mer at two sides of the monomer become one big 
$(k+k'+2)$-mer.

Therefore, any change in the coverage results from monomer movement.
Consequently, short times behavior is dominated firstly only by the 
monomer concentration. In latter times, monomers formed by the 
destruction of dimers contribute to the coverage as well, and thus 
the dimers, trimers, ... concentrations also play a role. In the 
short time regime, the dynamics of the $k$-mers concentration $c_k$
is dominated by the transitions of a $k$-mer to a $(k\pm 1)$-mer. 
The other process of a unification of a $k$-mer, a monomer and a 
$k'$-mer to a long $(k+k'+2)$-mer can be neglected to first orders 
in time, since it generates only $4$-mers or longer chains, and these 
do not contribute to the density up to fifth order. We thus have the 
following rate equations
\begin{eqnarray}
\dot c_1 & = & \epsilon (-2c_1 + c_2)  \nonumber \\
\dot c_k & = & \epsilon (c_{k-1}-2c_k+c_{k+1}), \quad\quad k>1
\end{eqnarray}
with the initial conditions (\ref{initial}).
Succesive approximations can be obtained by truncating the equation 
system after $n$ equations, fixing $c_{n+1}$ at its initial value. 
The result for $c_1$ is then exact for $n$ orders, and the coverage,
which is given by
\begin{equation}
\Delta\rho (t) = \epsilon\int_0^t c_1(t)dt
\end{equation}
is exact for $n+1$ orders. We thus get the first $4$ approximations
\begin{equation}
\Delta^{(4)}\rho(t)=c_1^0(\epsilon t) + 
(-2c_1^0+c_2^0)\frac{(\epsilon t)^2}{2} + 
(5c_1^0-4c_2^0+c_3^0)\frac{(\epsilon t)^3}{6} + 
(-14c_1^0+14c_2^0-6c_3^0+c_4^0)\frac{(\epsilon t)^4}{24} + O(t^5)
\end{equation} 
Figure 1 presents a comparison of the first $4$ approximations
with the results obtained from a numerical simulation using
a $N=256K$ lattice. Clearly, the fourth order expansion approximate
the real curve up to $\epsilon t=1$.

In order to study the long time behavior, we change our point of view.
We term each two adjecent empty sites between successive $k$-mers a
``hole''. Diffusion of particles at the edges of the monomers is
equivalent to the diffusion of these holes \cite{p}. 
At the course of the
hole diffusion, when two such holes are on adjacent sites, a space for
a new particle is formed, and after its deposition the two holes 
annihilate. We thus see that our model is equivalent to the model 
of $N$ random walkers 
on a lattice which annihilate each other when joined. This model is 
well known and was used especially to describe the dynamics of chemical
reactions of the type $A+A\to {\rm inert}$ \cite{lu,spouge,ba}. 
In what follows we apply
the analytical treatment developed for the reaction-diffusion problem
to our model, and obtain an asymptotic series for the density.
The two series, the asymptotic one and the previous short times 
expansion, describe the entire time regime very well, as can be 
shown by numerical results.

We wish to map our
model to a standard model of annihilating random walks for which
a rigorous result is known. We note that there are some differences
between this standard model and the diffusing holes: (a) when a 
particle at the edge of a $k-$mer moves, the hole changes its position 
by {\it two} lattice sites. Thus, the diffusion constant is four
times larger. (b) The distance between adjacent holes is always
an odd number of sites. In particular, in the initial configuration
the distances are odd. (c) The annihilation process occurs whenever 
the distance between the holes is one site and not when they are on the
same site as in the standard model. Accordingly, in the initial state
the minimal distance between successive $k-$mers is three sites.
However, one expects these two differences to have no effect for long
times for which the behavior is dominated by holes far from each other.
This is confirmed by our numerical results.

Thus we consider a model of random walks (RWs), originally distributed 
randomly on the lattice, with density 
\begin{equation}
\tr\equiv\tr(0)=2(1-\rho_r)=e^{-2},
\end{equation}
where $\rho_r$ is the jamming limit density. The RWs moves
with probability $\epsilon$ two sites to right or left. When two such
RWs join, they are annilihated. This model was solved exactly
\cite{spouge,ba}, and we here follow the derivation given 
by Spouge \cite{spouge}. Define $\beta_k(n)$
to be the probability of the $k$th RW to be at a distance $n$ from
the origin
\begin{equation}
\beta_k(n) = \tr^k(1-\tr)^{n-k}\left ( \frac{n-1}{k-1}\right ).
\end{equation}
$\beta_-(n)$ is then defined by
\begin{equation}
\beta_-(n):=\delta_{n,0}+2\sum_{k=1}^\infty (-)^k\beta_k(n)=
\delta_{n,0}-2\tr(1-2\tr)^{n-1}.
\end{equation}
One also defines $a(t;n)$ which is the probability that two RW 
whose original distance was $n$ have meet until time $t$.
In our model
\begin{equation}
a(t;n)=\I_n(4Dt)+2\sum_{k=n+1}^\infty \I_k(4Dt)
\end{equation}
where $\I_n(x):=e^{-x}I_n(x)$, $I_n(x)$ is the modified Bessel 
function of integer order, and $D$ is the diffusion constant of 
the particles.

Given the above definitions, Spouge's main result is
\begin{equation}
\tr(t) = \tr \sum_n a(t;n)\beta_-{n}
\end{equation}
Substituting the above expressions for our model we get
\begin{eqnarray}
\tr(t)/\tr &=& 1 - 2\tr \sum_{n=0}^\infty (1-2\tr)^n[\I_{n+1}(4Dt)+
2\sum_{k=n+2}^\infty \I_k(4Dt)] \nonumber \\
& = & 1 - 2\tr\sum_{k=1}^\infty \I_k(4Dt)[q^{k-1}+2q^{k-2}+ ... +
2q^0] \nonumber \\
& = & 1 - \sum_{k=1}^\infty (2-q^{k-1}-q^k)\I_k(4Dt)\nonumber \\
& = & \I_0(4Dt) + (1+q)\sum_{k=1}^\infty q^{k-1}\I_k(4Dt)
\label{sum}
\end{eqnarray}
where $q=1-2\tr$. A similar expression was given by Balding
{\it et al} \cite{ba}.
The particle density is given in terms of the hole density through
the relation $\rho = (1-\tr)/2$, and thus one obtains for the 
difference between the density and the maximal, closest packing, 
density
\begin{equation}
\rho_{cp} - \rho(t) = \tr(t)/2 = \frac{e^{-2}}{2}[\I_0(4Dt) + 
(1+q)\sum_{k=1}^\infty q^{k-1}\I_k(4Dt)]
\end{equation}

The diffusion constant is determined easily through the relation
$<r^2>=2Dt$ resulting in $D=2\epsilon$. Now, the asymptotic behavior 
follows from the known asymptotics of the Bessel functions
\cite{sa}. 
\begin{equation}
\I_k(z)= \frac{1}{\sqrt{2\pi z}}[1 - \frac{\mu-1}{8z}+
\frac{(\mu-1)(\mu-9)}{2!(8z)^2} - ...],\quad\quad\quad\quad \mu=4k^2
\end{equation}
To the first order, all the $\I$s are identical ($k$ independent) 
and one has
\begin{equation}
\label{a1}
\rho_{cp}-\rho(t)\sim \frac{\tr}{2\sqrt{16\pi\epsilon t}}[1+\frac{1+q}
{1-q}]=\frac{1}{8\sqrt{\pi\epsilon t}}=\frac{0.0705....}
{\sqrt{\epsilon t}}.
\end{equation}
In a similar way, next orders can be extracted. For example, the next 
correction is
\begin{equation}
\rho_{cp}-\rho(t)=\frac{1}{8\sqrt{\pi\epsilon t}}-\frac{a_2}
{(\epsilon t)^{3/2}}+O((\epsilon t)^{-5/2}).
\end{equation}
where
\begin{equation}
a_2=\frac{2e^4-4e^2+1}{512\sqrt{\pi}}=0.08886... 
\end{equation}

Figure 2 presents a comparison of the asymptotic leading order 
(\ref{a1}) and the series (\ref{sum}) with numerical results obtained 
from a lattice of $N=1.2M$ sites. One sees that the whole series
fits the results even for small values of $\epsilon t$ down to
$\epsilon t=0.1$.

In summary, it has been shown that the 1D deposition-diffusion
process leads to full coverage. The short time dynamics is 
determined by the temporal momnomer concentration. A fourth order
expansion is given, valid up to $\epsilon t =1$. The long time
kinetics is dominated by the attachment of two relatively long
$k$-mers which forms one long $k'$-mer. This process is equivalent to
the dynamics of a reaction-diffusion process, or to the probability
of a RW to return to the origin. Thus, the asymptotic approach
of the density to its saturated value in 1D is $O(1/\sqrt{t})$.
We derive an asymptotic series, based on this equivalency, valid 
for the intermediate and long time regimes ($\epsilon t > 0.1$).

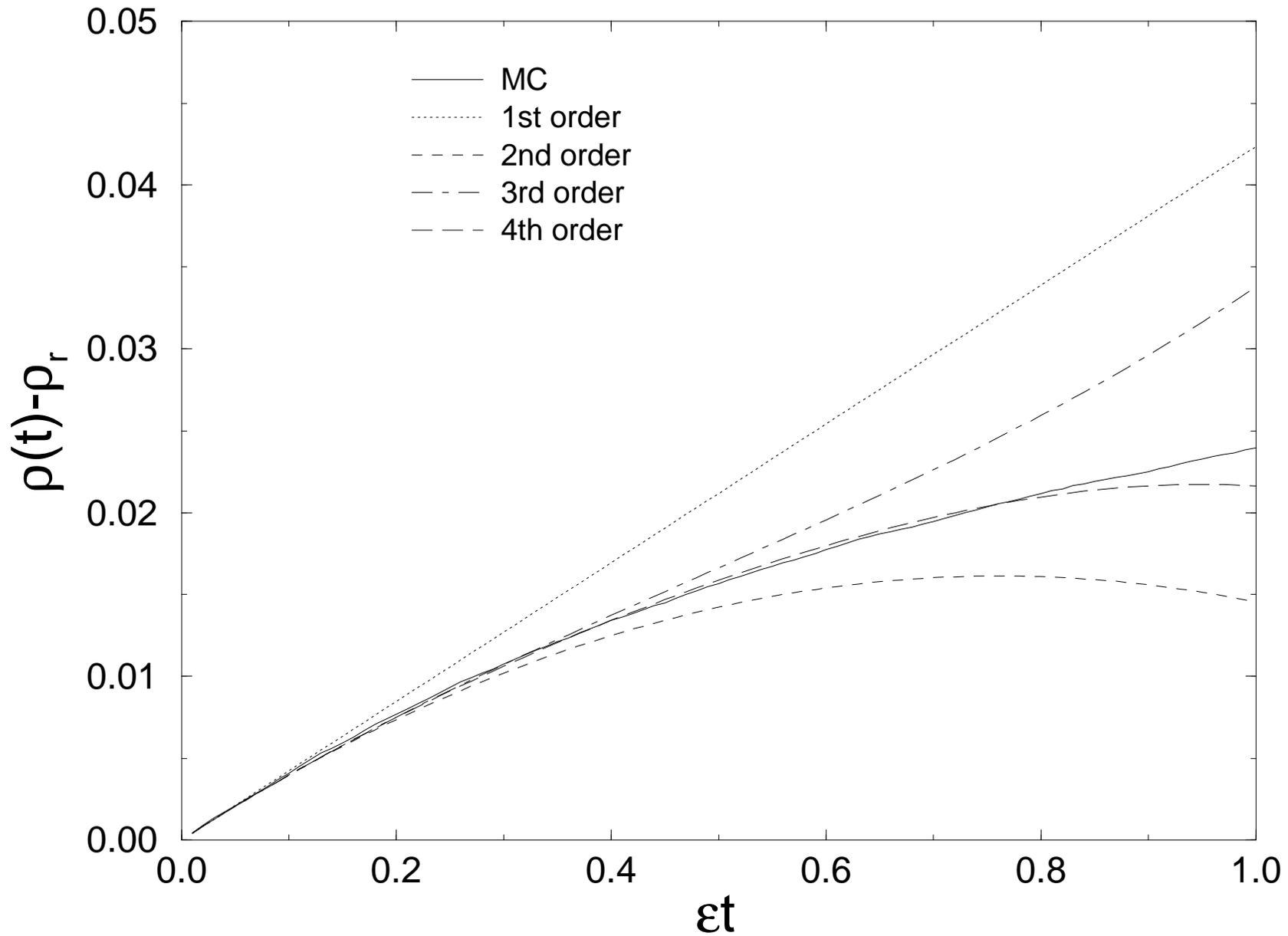
\begin{figure}
\caption{Numerical results for the coverage for short times, plotted
vs. $\epsilon t$, compared to the first four short-times 
approximations.}
\end{figure}

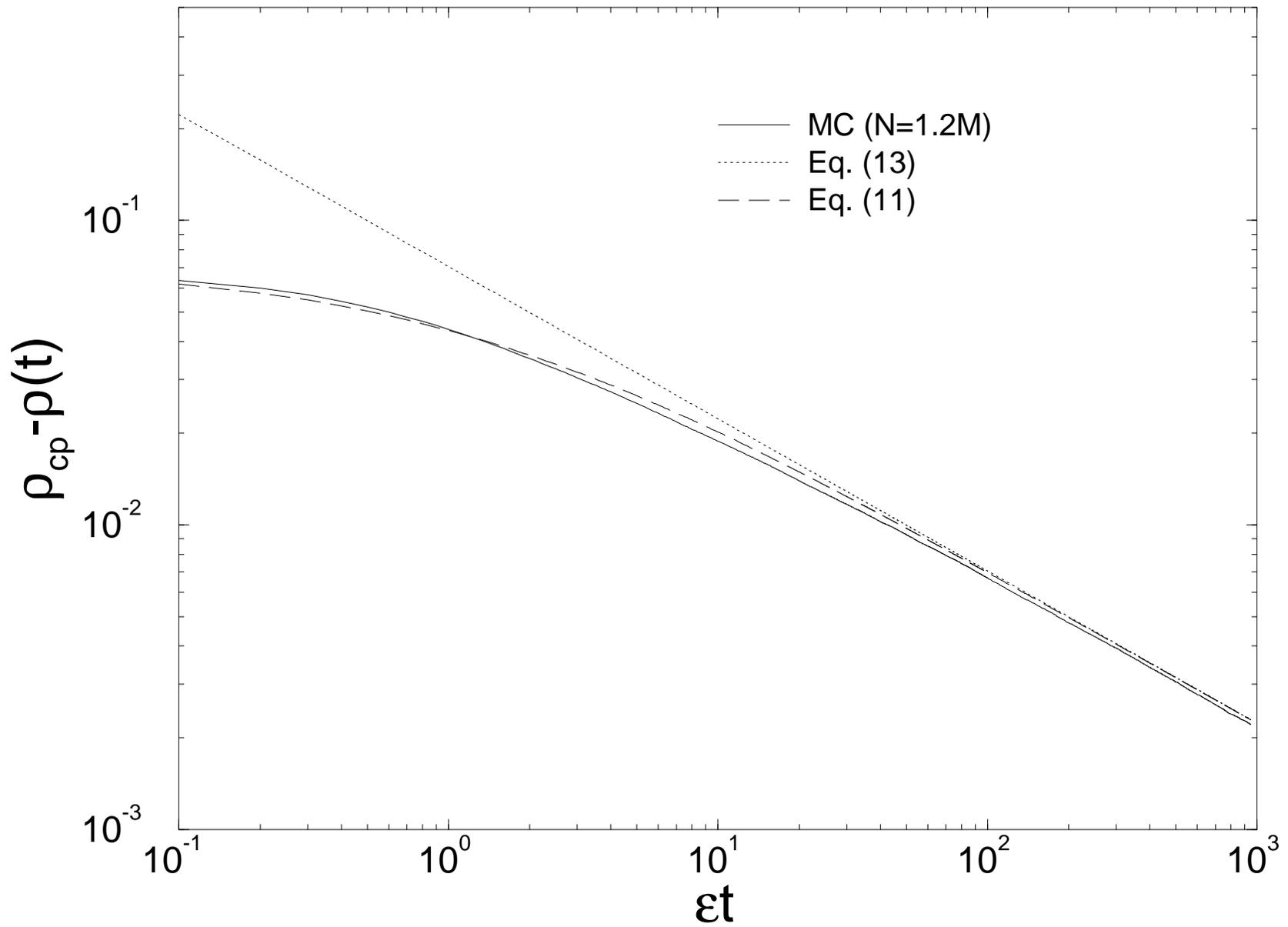
\begin{figure}
\caption{Numerical results for the coverage for long times, plotted
vs. $\epsilon t$, compared to the asymptotic leading order, and the
Bessel functions' sum 
.}
\end{figure}

\end{document}